\documentclass[sigconf,authorversion,nonacm]{acmart}

\acmConference[CCSW'24]{ACM Cloud Computing Security Workshop}{10/18/2024}{Salt Lake City, U.S.A.}

\copyrightyear{2024}
\acmYear{2024}
\setcopyright{acmlicensed}\acmConference[CCSW '24]{Proceedings of the 2024 Cloud Computing Security Workshop }{October 14--18, 2024}{Salt Lake City, UT, USA}
\acmBooktitle{Proceedings of the 2024 Cloud Computing Security Workshop (CCSW '24), October 14--18, 2024, Salt Lake City, UT, USA}
\acmDOI{10.1145/3689938.3694781}
\acmISBN{979-8-4007-1234-0/24/10}

\usepackage{amsmath}
\usepackage{opera}
\usepackage{balance}

\usepackage{multirow}
\usepackage{graphicx}

\usepackage{enumitem}

\newcommand{\ebpf}{eBPF\xspace}
\newcommand{\system}{SafeBPF\xspace}

\newcommand*\circled[1]{{\raisebox{.5pt}{\textcircled{\raisebox{-.9pt} {#1}}}}}

\definecolor{lightblue}{HTML}{2E75B6}
\definecolor{peachyorange}{HTML}{C55A11}
\newcommand{\tagkernel}{\textcolor{peachyorange}{\circled{b}}}
\newcommand{\tagsandbox}{\textcolor{lightblue}{\circled{a}}\xspace}

\definecolor{darkgreen}{HTML}{013220}

\definecolor{programcolor}{HTML}{87CEEB}
\definecolor{contextcolor}{HTML}{FF6347}
\definecolor{sandboxcolor}{HTML}{D8BFD8}
\definecolor{accesscolor}{HTML}{FFA500}
\definecolor{taggingcolor}{HTML}{8DB600}

\definecolor{vanillacolor}{HTML}{C71585}
\definecolor{mtecolor}{HTML}{006400}

\definecolor{verifiercolor}{HTML}{00008B}

\newcommand{\program}{\textcolor{programcolor}{$\blacksquare$}~\textit{\ebpf~Program}\xspace}
\newcommand{\context}{\textcolor{contextcolor}{$\blacksquare$}~\textit{Context~Synchronization}\xspace}
\newcommand{\sandbox}{\textcolor{sandboxcolor}{$\blacksquare$}~\textit{Sandbox~Management}\xspace}
\newcommand{\access}{\textcolor{accesscolor}{$\blacksquare$}~\textit{Access~Checks}\xspace}
\newcommand{\tagging}{\textcolor{taggingcolor}{$\blacksquare$}~\textit{Object Tagging}\xspace}

\newcommand{\mte}{\textcolor{mtecolor}{$\blacksquare$}~\textit{mte}\xspace}

\newcommand{\vanilla}{\textcolor{vanillacolor}{$\blacksquare$}~\textit{vanilla}\xspace}

\begin{document}

\date{}

\title{\system: Hardware-assisted Defense-in-depth for \ebpf Kernel Extensions}

\author{Soo Yee Lim}
\email{sooyee@cs.ubc.ca}
\orcid{0000-0002-3418-4982}
\affiliation{%
	\institution{University of British Columbia}
	\city{Vancouver}
	\state{British Columbia}
	\country{Canada}
}

\author{Tanya Prasad}
\email{tanyapsd@cs.ubc.ca}
\orcid{0009-0000-5378-1857}
\affiliation{%
	\institution{University of British Columbia}
	\city{Vancouver}
	\state{British Columbia}
	\country{Canada}
}

\author{Xueyuan Han}
\email{vanbasm@wfu.edu}
\orcid{0000-0003-1374-153X}
\affiliation{%
\institution{Wake Forest University}
\city{Winston-Salem}
\state{North Carolina}
\country{USA}
}

\author{Thomas Pasquier}
\email{tfjmp@cs.ubc.ca}
\orcid{0000-0001-6876-1306}
\affiliation{%
	\institution{University of British Columbia}
	\city{Vancouver}
	\state{British Columbia}
	\country{Canada}
}

\begin{abstract}
	The \ebpf framework enables
execution 
of user-provided code
in the Linux kernel.
In the last few years,
a large ecosystem of cloud services has leveraged \ebpf to enhance container security, system observability, and network management.
Meanwhile,
incessant discoveries
of memory safety vulnerabilities
have left the systems community
with no choice
but to disallow
unprivileged \ebpf programs,
which unfortunately limits
\ebpf use 
to only privileged users.
To improve run-time safety
of the framework,
we introduce \system,
a general design 
that isolates 
\ebpf programs 
from the rest of the kernel 
to prevent memory safety vulnerabilities 
from being exploited.
We present
a pure software
implementation
using a
Software-based Fault Isolation (SFI) approach
and a hardware-assisted
implementation
that leverages 
ARM's Memory Tagging Extension (MTE).
We show that
\system
incurs up to 4\% overhead 
on macrobenchmarks
while achieving desired security properties.

\noindgras{Note:} This is a preprint of the paper accepted at the 2024 ACM Cloud Computing Security Workshop~\cite{soo2024safebpf}.

\end{abstract}

\maketitle

\section{Introduction}
\label{sec:introduction}
In recent years, 
we have been witnessing an increasing uptake,
both in academia and industry,
of using 
the extended Berkeley Packet Filter (\ebpf) 
to customize in-kernel behavior.
The \ebpf framework is designed 
to enable safe extension of the Linux kernel 
without modifying the kernel source code.
Prior work~\cite{lim2021secure, belair2021snappy, sekareaudit, shahinfar2023automatic, zhong2021bpf, yang2023lambda, park2022application,cao2024atc}
has demonstrated
success of leveraging \ebpf
in a variety of use cases,
ranging from
packet forwarding
to balance network traffic load~\cite{katran}
and network monitoring
to secure and troubleshoot
communications
in a microservices architecture~\cite{cilium,pixie},
to application-specific
scheduling~\cite{kaffes2021syrup}, prefetching~\cite{cao2024atc} and
page cache management~\cite{lee2023p2cache}.
Companies like Tigera~\cite{tigera} and Cilium~\cite{cilium} have capitalized on \ebpf to deliver container security, networking, and observability solutions for modern cloud computing environments.
However, unprivileged containers are often unable to leverage these \ebpf features as they require privileges not granted to untrusted containers.

Indeed, the safety of the \ebpf framework 
relies primarily on \emph{statically} verifying 
an \ebpf program 
before it is allowed to run
in the kernel.
Unfortunately, 
recent work~\cite{lim2023ebpf,jia2023kernel} 
has shown that 
static verification alone is insufficient
to prevent
malicious \ebpf programs
from %
accessing arbitrary kernel
 memory~\cite{cve-2021-34866, cve-2020-8835, cve-2021-31440, cve-2021-3490, cve-2021-4204, cve-2022-23222, cve-2023-2163,cve-2021-33200}. %
Consequently, most kernel distributions 
disable the use of 
\ebpf %
by unprivileged users~\cite{unpriv_ubuntu,unpriv_suse}. %
This significantly
limits its adoption
and sometimes even
encourages
deliberate, unsafe practices,
thereby defeating the purpose of
disallowing unprivileged uses
in the first place.
For example,
practitioners
are often interested in
running \ebpf programs in unprivileged containers.
With privileged-only
execution restriction in place,
the general consensus
is to circumvent this restriction
using privileged processes,
rather than finding safer alternatives.
While this is only
an anecdotal example,
there is no denying that
security,
when done at the expense of usability
(or convenience),
often loses its priority.

If we improved the \ebpf verifier,
then this could potentially 
guarantee the safety
of unprivileged \ebpf programs and
therefore
address the concerns
that (rightfully) hinder \ebpf's
wide deployment.
However,
latest work~\cite{jia2023kernel} has shown
that such efforts,
like fuzzing~\cite{fuzzing-ebpf-1,fuzzing-ebpf-2}
and formal verification~\cite{vishwanathan2023verifying,vishwanathan2022sound,bhat2022formal},
are insufficient,
due %
to the constantly increasing complexity 
of the verifier.
A major overhaul of
completely retiring the current verifier
and using instead
a memory-safe language like Rust~\cite{jia2023kernel}
only shifts the problem
from the verifier
to the external Rust toolchain.

We introduce \system,
a \emph{dynamic sandboxing} approach
that works alongside the verifier to improve \ebpf security by
isolating \ebpf programs
from the rest of the kernel.
Using a combination of 
\emph{software-based fault isolation}
and \emph{hardware-implemented memory tagging}
techniques,
\system confines 
all memory accesses of an \ebpf program
to a well-defined sandbox,
thus preventing run-time violations of spatial memory safety,
even if the vulnerabilities
that lead to these violations
bypass static verifier checks.
Our evaluation shows that 
\system can effectively
prevent memory bugs 
missed by the \ebpf verifier
while incurring at most 4\% performance overhead 
on marcobenchmarks.

\noindgras{Contributions}
\begin{itemize}[leftmargin=*]
\setlength\itemsep{0em}	
	\item We propose a new execution environment 
        for \ebpf extensions (\autoref{sec:design}) 
        and explore different mechanisms to dynamically enforce spatial memory safety (\autoref{sec:implementation}).
    \item We systematically evaluate \system and show that it introduces low run-time performance overhead 
    while improving \ebpf security (\autoref{sec:evaluation} and \autoref{sec:discussion}).
    \item To the best of our knowledge, 
    we are the first to directly compare software-based isolation 
    and ARM's MTE as alternative mechanisms 
    to achieve in-kernel isolation (\autoref{sec:evaluation:performance:micro}).
    \item We make our \system implementation
        and the corresponding patches to the Linux kernel
        publicly available 
        for the community to expand upon.
        We also make the materials to reproduce the evaluation
        available online.\footnote{Software artifacts (under GPLv2 license) and materials to reproduce the evaluation are available online at \url{https://s00y33.github.io/publication/safebpf/SafeBPF.patch}.
       	We provide kernel patches for software-based isolation on both the x86 and ARM architecture, and hardware-assisted isolation on ARM.}
\end{itemize}

\section{Background \& Motivation}
\label{sec:background}
The extended Berkeley Packet Filter (eBPF) 
is a Linux framework 
that enables users to extend the kernel's capabilities 
without modifying its source code
or loading additional kernel modules.
The original %
BPF
was designed for 
packet filtering;
its functionality
has since been extended 
as the underlying technology
to drive
a wide array of applications
in areas such as
performance monitoring~\cite{pixie}, 
system tracing~\cite{bcc_tracing}, 
load balancing~\cite{katran}, 
and security~\cite{falco}.

Kernel extensions, 
including those enabled by \ebpf, 
can pose security risks to a system.
Due to a lack of isolation 
in a monolithic operating system (OS), 
a vulnerability in a kernel extension 
grants an attacker full access 
to the rest of the OS 
with which the extension shares an address space.
While \ebpf is supposed to ensure safety 
through static verification,
\emph{run-time safety} remains an open problem
(\autoref{sec:background:runtime_safety}).

\begin{figure}[!t]
	\centering
	\includegraphics[width=\columnwidth]{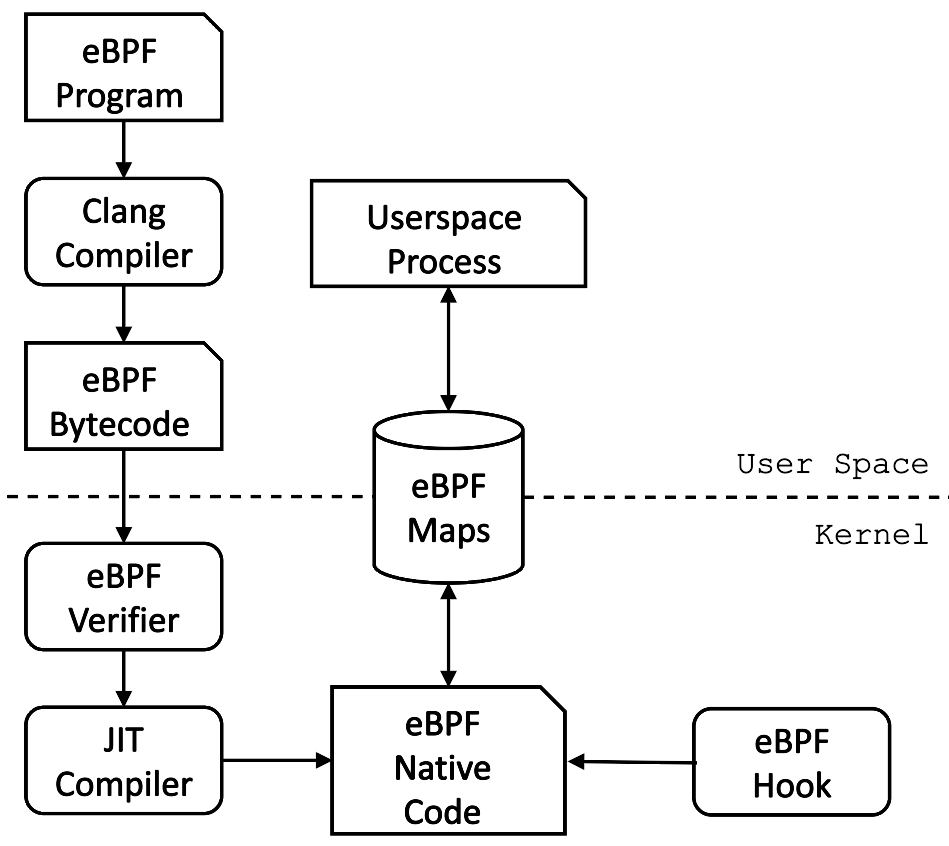}
	\caption{An overview of eBPF workflow.}
	\label{fig:ebpf_workflow}
\end{figure}

\subsection{An Overview of \ebpf}
\ebpf programs can be written in many high-level programming languages, such as C and Rust.
\autoref{fig:ebpf_workflow} shows the general workflow of an \ebpf program written in C.
First, an \ebpf program is compiled 
by Clang/LLVM to generate an ELF binary that contains architecture-independent \ebpf bytecode.
An \ebpf ELF loader (\eg libbpf) then parses the ELF binary and does the heavy lifting of preparing and loading the \ebpf program into the Linux kernel.
During loading,
the kernel's \ebpf verifier statically checks the safety of the \ebpf bytecode.
A Just-In-Time (JIT) compiler then translates the verified \ebpf bytecode into native machine code, so the \ebpf program can run as efficiently as natively compiled kernel code.
Once the JIT compilation is completed, the \ebpf program is marked as read-only to prevent any corruptions throughout its lifetime.
Finally, the program is attached to its designated kernel hook point where 
it gets triggered and executed at run time.
An \ebpf program can interact with userspace processes via special
shared data structures called \ebpf maps.
It is also restricted to a well-defined interface,
i.e., the \ebpf helper functions,
to interact with the kernel.
Depending on the program type,
an \ebpf program
can access
only a subset of the helper functions.

\subsection{The \ebpf Verifier}
\label{sec:background:verifier}

The \ebpf verifier performs a two-pass 
static verification of the \ebpf bytecode.
The first pass conducts a depth-first search
to reject
\ebpf programs 
that contain unreachable instructions, 
unbounded loops, 
or out-of-bounds jumps.
It also rejects
\ebpf programs
that are too large
(exceeding $4,096$
instructions for unprivileged programs
and $1$M for privileged ones)
for the verifier
to perform static analysis.
Some of these restrictions are incompatible 
with compiler optimizations~\cite{verifier-llvm-clash};
as such, 
correctly compiled \ebpf programs 
can be rejected by the \ebpf verifier.
The proposal to integrate the \ebpf verifier 
into the compiler could potentially 
address this issue~\cite{verifier-in-llvm},
but maintaining such a compiler
is difficult
since the \ebpf framework
is constantly evolving at a fast pace
(see \autoref{sec:background:panacea}).

The second pass %
simulates program execution,
tracking its state (i.e., registers and stack)
changes
to catch unsafe operations (\eg out-of-bounds accesses).
On entry to each instruction, 
each register is assigned a type;
the simulation of instruction execution
changes the types of the registers 
depending on instruction semantics.
For example, 
the \ebpf verifier forbids
an instruction from adding
two pointers.
Doing so would result 
in a register state 
of type \texttt{SCALAR\_VALUE},
indicating a non-pointer
register value.
Any subsequent instruction
attempting to access 
the register value
as a pointer
would then be rejected by the verifier.%

\begin{figure}[!t]
\centering
\includegraphics[width=\columnwidth]{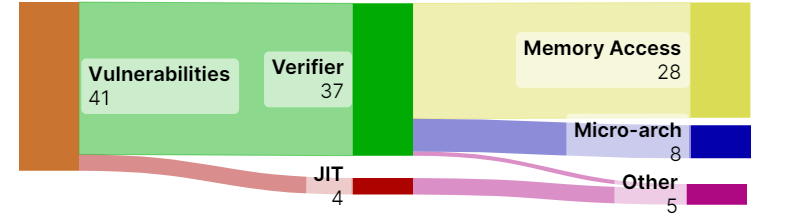}
\caption{A summary of the types of CVEs reported in each component of \ebpf from 2010 to 2023.}
\label{fig:ebpf-vuln}
\end{figure}

\subsection{\ebpf Lacks Run-time Memory Safety}
\label{sec:background:runtime_safety}

Vulnerabilities in the \ebpf verifier often lead to bypasses of static security checks
by the verifier,
many of which have already
been recorded in the
Common Vulnerabilities and Exposures (CVE) database.
Those bugs are related to 
\emph{speculative execution} 
and \emph{memory safety}.
While general kernel hardening~\cite{kernel-speculative}
and specific \ebpf changes~\cite{bpf-speculative}
have, for the most part,
addressed the former concern,
the latter,
which comprises 
the majority of discovered vulnerabilities
($28$ out of $41$, 
approximately $68$\%, 
see~\autoref{fig:ebpf-vuln}),
remains a critical issue.
Memory safety
can be enforced by
\emph{dynamic} security checks;
unfortunately, 
these checks are absent,
since
verified \ebpf programs run in the same address space as the rest of the kernel and are assumed 
to be trusted and safe.
Adversaries exploiting
memory-unsafe \ebpf programs
could therefore
gain full access to kernel memory at run time.
For example, 
CVE-2021-3490 is an out-of-bounds access vulnerability due to a bounds tracking bug 
in the \ebpf verifier's 
32-bit arithmetic and logic unit. 
This vulnerability grants unprivileged adversaries 
arbitrary read and write access to kernel memory,
which enables them 
to achieve privilege escalation
by overwriting the \texttt{cred} structure.

As eBPF continues
to grow in complexity,
with new features
such as \ebpf tokens~\cite{bpf_token} 
and \ebpf exceptions~\cite{bpf_exception}
being regularly incorporated into 
the mainline Linux kernel,
we can expect only \emph{more} vulnerabilities
to be reported in the future.
More concerningly,
a recent study~\cite{alexopoulos2022long}
shows that
vulnerabilities remain in the kernel 
for an average of $1,800$ days 
before being addressed.
The volume and
duration of
vulnerabilities
inevitably puts 
\emph{run-time} safety of \ebpf extensions 
at a high risk.
Not surprisingly, 
unprivileged \ebpf is %
by default
turned off on most Linux distributions~\cite{unpriv_ubuntu,unpriv_suse}.

\subsection{Improving the \ebpf Verifier Is No Panacea}
\label{sec:background:panacea}

\begin{figure}[!t]
	\centering
	\includegraphics[width=\columnwidth]{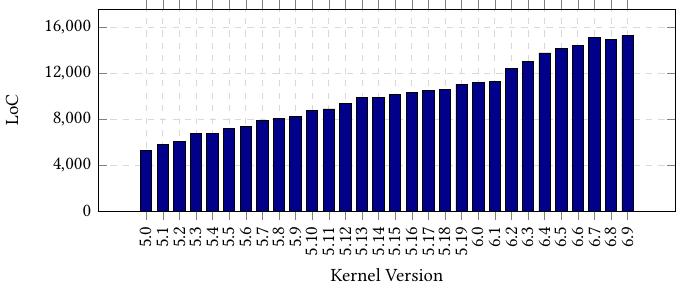}
	\caption{The evolution of the \ebpf verifier's size in source lines of code (SLOC) from v5.0 in March 2019 ($5,245$ SLOC) to v6.9 in May 2024 ($15,274$ SLOC).}
	\label{fig:verifier-size}
\end{figure}

Many \ebpf hardening solutions focus on improving the \ebpf verifier.
One approach is to formally verify the soundness of the verifier~\cite{vishwanathan2023verifying,vishwanathan2022sound,bhat2022formal}.
While this can eliminate implementation bugs, 
existing solutions verify only parts of the verifier, 
such as the tristate numbers (tnums) abstract domain~\cite{vishwanathan2022sound} and range analysis~\cite{vishwanathan2023verifying,bhat2022formal}.
It is in fact challenging 
to extend formal verification 
to the entire verifier, because
1) the verifier itself 
has no formal specification~\cite{gershuni2019simple};
2) its size has increased significantly 
to verify the safety of 
the growing set of \ebpf features 
(\autoref{sec:background:runtime_safety}),
which complicates formal verification
and makes it hard to scale;
and 3) emerging compiler features work directly against the verifier~\cite{verifier-llvm-clash}.
As we can see in~\autoref{fig:verifier-size},
the size of the \ebpf verifier 
has \emph{more than doubled} in the past four years,
creating ample opportunities
for attackers to find bugs 
to bypass static checks
and weaponize \ebpf programs.

Jia \etal~\cite{jia2023kernel}
propose to replace the \ebpf verifier
with the Rust compiler.
Rust is a memory-safe programming language 
that leverages the type system 
and an ownership model 
to eliminate memory safety bugs at compile time.
However, the Rust ecosystem is large and complex 
(approximately $1.6$M lines of Rust code). 
Vulnerabilities in Rust~\cite{cve_rust} 
lead us back to the same problem 
with the \ebpf verifier:
\emph{static checks alone 
cannot ensure run-time memory safety}.
Moreover, 
since the proposed Rust-based \ebpf design 
completely retires the \ebpf verifier, 
it delegates the role of 
authorizing safe \ebpf programs 
to a trusted third party in userspace.
As a result, 
the kernel can load only \ebpf programs 
signed by those parties, 
as it can no longer independently verify them.
This design limits \ebpf usage, 
sacrificing kernel extensibility for security.
We note also that 
in a similar space, 
the driver signature scheme 
has been exploited 
by attackers~\cite{driver-sign,cve-2021-35039}.

\subsection{Our Proposed Approach}

To bridge the security gap at run time, 
we propose to dynamically isolate \ebpf programs,
in addition to static verification. 
Since the majority of the \ebpf verifier's vulnerabilities 
stem from bounds checking, 
we explore both software- and hardware-based 
isolation techniques 
to prevent run-time out-of-bounds kernel memory access. 
In particular, 
\system confines all memory accesses 
to within the \ebpf sandbox,
where all \ebpf data resides.
We show that \ebpf isolation is feasible 
with software-based fault isolation, 
and its performance overhead can be reduced when using ARM Memory Tagging Extension (MTE).
We discuss the applicability of our approach 
to other CPU architectures in~\autoref{sec:discussion}.

\begin{figure*}[t]
	\centering
	\captionsetup{justification=centering,margin=2cm}
	\includegraphics[width=\textwidth]{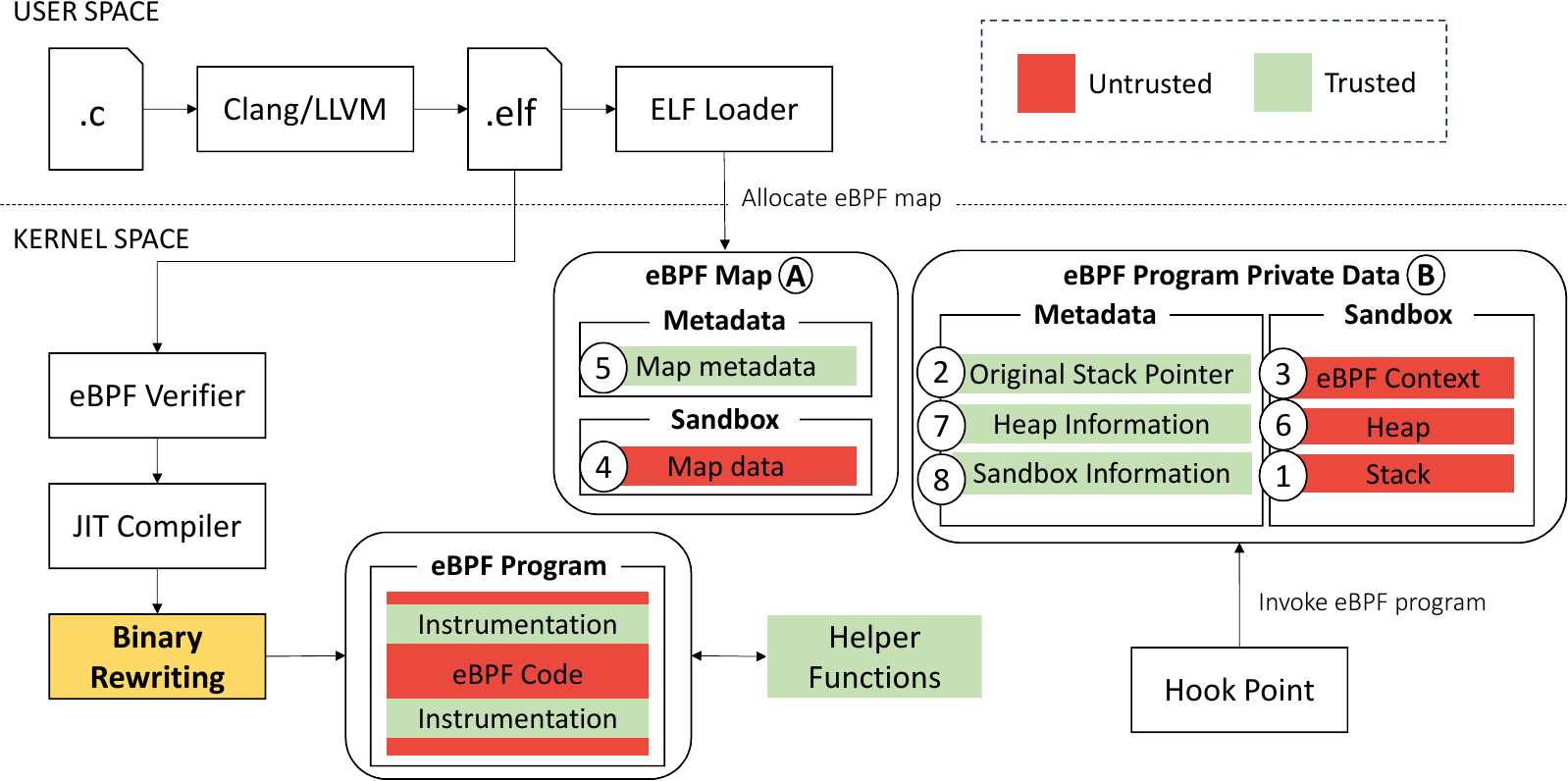}
	\caption{An illustration of the \system design.}
	\label{fig:design}
\end{figure*}

\section{Threat Model}
\label{sec:threat}
\system targets \emph{spatial memory safety}.
In particular, we assume 
adversaries 
running \emph{unprivileged} \ebpf programs
without root access,
thus unable to load kernel modules 
or modify kernel code.
However, 
they can exploit vulnerabilities 
in the \ebpf verifier 
to bypass memory access checks, 
therefore gaining arbitrary read or write access 
to kernel memory.
We assume a W$\oplus$X (write xor execute) enabled system,
so attackers cannot overwrite any executable pages.

Our trusted computing base includes the OS kernel (excluding the \ebpf verifier) 
and \system.
More specifically,
as shown in~\autoref{fig:design},
we assume that
all the data in an \ebpf program, 
including both private 
(\eg stack and context) 
and shared data (\eg \ebpf maps),
are untrusted.
We also assume that
\system's instrumentation
and its own data,
which are stored outside of the sandbox
(see details in~\autoref{sec:design}),
are trusted. 

\system instruments 
the final output of \ebpf's JIT compilation;
therefore,
it does not rely on
the correctness of the JIT compiler or the verifier.
However, 
since we require \ebpf programs to be \emph{compiled}, 
we disable the legacy \ebpf interpretation feature~\cite{bpf-jit-always-on-misc}
to enforce JIT compilation.
Note that \ebpf interpretation is insecure 
due to its own set of vulnerabilities~\cite{jin2023epf}
and has already been 
disabled by default on
the x86 and ARM architectures 
for most distributions.
Securing \ebpf interpretation is 
beyond the scope of this paper.

Like in prior kernel isolation work~\cite{swift2003improving, erlingsson2006xfi, ganapathy2007microdrivers, renzelmann2009decaf, castro2009fast, mao2011software,nikolaev2013virtuos,narayanan2019lxds,narayanan2020lightweight,mckee2022preventing}, 
side-channel attacks are orthogonal and thus out of scope (see further discussion in~\autoref{sec:discussion}).

\section{Sandbox Design}
\label{sec:design}
Isolating an \ebpf program involves
1) creating a sandbox and 2) enforcing isolation.
In this section,
we describe at a high level
how \system
constructs a sandbox
for an \ebpf program
(\autoref{fig:design}).
We then detail different mechanisms
to enforce isolation in~\autoref{sec:implementation}
and compare their performance in~\autoref{sec:evaluation}.

We emphasize that
\system is \emph{not} intended to replace the \ebpf verifier;
in fact,
we still rely on
it
to perform
static checks
(\eg type checking the \ebpf helper functions)
and post-verification rewrites.
However,
in addition to these checks,
\system fortifies
\emph{defense-in-depth}
by dynamically isolating \ebpf programs
to ensure spatial memory safety at run time,
thereby protecting the kernel
from most known \ebpf vulnerabilities.

\subsection{Requirements}
\label{sec:design:requirements}
We design \system
with the following
first-order requirements in mind.
The remainder of~\autoref{sec:design}
and~\autoref{sec:implementation}
explain how we successfully
achieve these requirements.

\noindgras{Isolation.}
\system must isolate the kernel from \ebpf exploits.
It must prevent run-time violations
of spatial memory safety
and stop malicious \ebpf programs
from corrupting or leaking arbitrary kernel memory
to userspace.

\noindgras{Efficiency.}
Performance is crucial to \ebpf,
especially since certain \ebpf program types
are usually attached
to critical code paths in the kernel.
For example,
an XDP program
is attached to a network interface card
to process network packets.
\system only minimally
affects the overall performance of the kernel,
incurring at most
4\% overhead on the Apache benchmark.
We justify this reasonably
small performance cost
by the substantial security
\emph{and} usability benefit
to the \ebpf framework,
particularly since
unprivileged \ebpf programs are currently disallowed
in most Linux distributions
due to the risks
associated with spatial memory safety.

\noindgras{Portability.}
\system can run on any x86-64 and ARM64 platforms.
While \system leverages hardware extensions,
its design also supports
a fully software-based approach
when the required hardware feature is unavailable.
We carefully compare the overhead introduced by our software and hardware techniques in \autoref{sec:evaluation}.

\noindgras{Minimally Invasive.}
\system is seamlessly integrated
into the current \ebpf pipeline,
extending only what is necessary.
Experienced users can thus develop \ebpf programs
like they normally would,
and new users can still rely on
the existing documentation.
\system supports
existing \ebpf programs \emph{as-is}
and is entirely transparent to the end-user.

\subsection{Sandbox Construction}
\label{sec:design:isolation}

\system identifies
its sandbox's isolation boundary
without users' manual annotation.
Precisely defining
the isolation boundary
is important,
since it determines
what kernel data
is and is not
accessible to an \ebpf program
and therefore has
security implications.
Meanwhile,
it is also a challenging
problem,
often addressed
manually~\cite{witchel2005mondrix,swift2003improving,safedrive06,castro2009fast,mao2011software,narayanan2019lxds,narayanan2020lightweight,mckee2022preventing}
or semi-automatically~\cite{ganapathy_microdrivers_2008,renzelmann2009decaf,huang2022ksplit}
in prior
dynamic sandboxing work.
However, this is not necessary in \system
thanks to
the \ebpf framework's already
well-defined API and rules around the kinds of data
that different types of \ebpf programs
are allowed to access.
We use this information to define our isolation boundaries.

More specifically,
since \ebpf programs
reside in the same virtual address space as the kernel,
\system creates
a \emph{logically separated} sandbox
in the kernel address space
(\autoref{fig:design}).
\system further divides the sandbox
into different components
based on data types:
For a single \ebpf program instance,
one component stores
the program's private data,
\ie its stack, heap, and context;
every \ebpf map
also has a separate component.
Although \ebpf programs are allowed to access other kernel objects by obtaining their pointers from helper functions (\eg the \texttt{bpf\_sk\_lookup\_tcp} helper function returns a TCP socket pointer to the \ebpf program), \system does not protect accesses to these data as these features are not designed to be used by unprivileged users.
In fact, so-called ``offensive'' \ebpf features (\eg the helper function \texttt{bpf\_probe\_read\_user} that allows \ebpf programs to read the memory of any process) should not be considered for container use because they can be exploited to break resource isolation in OS-level virtualization~\cite{he2023cross}.

Each component
is associated with a \emph{metadata} region
that is inaccessible
to \ebpf programs;
metadata are
used by the kernel
(and \system)
to manage the associated data objects.
We compartmentalize
the sandbox
this way
for two reasons.
First, \ebpf maps have
a different lifetime
than that of
program-private data.
While an \ebpf map is freed
only when no reference to it exists,
data of the other types
will immediately go out of scope
upon a program's exit.
By isolating them in separate components,
\system can manage
the life-cycle
of a sandbox's components
based on the lifetime of the data they isolate.
Second,
since \ebpf maps can be shared
among multiple \ebpf programs
while the other data types cannot,
separating them
makes it easy for \system
to allow shared access to maps
while restricting the other components
to only their corresponding \ebpf program instances.
We %
detail
how \system constructs
the sandbox
next.

\subsubsection{\ebpf Maps}
\label{sec:construction:maps}

\ebpf maps (\circled{A} in~\autoref{fig:design})
are memory regions shared
among multiple \ebpf programs and
between an \ebpf program
and a userspace application.
They must be accessed
through dedicated helper functions.
The \ebpf framework implements
various types of maps
with different semantics
(\eg hash, array, and bloom filter).
Along with map data,
the framework
allocates metadata (\circled{5} in~\autoref{fig:design}),
such as the reference counter
and synchronization primitives,
to manage a map and its specific semantic.
Regardless of the
type of an \ebpf map,
\system applies
the same isolation principle.
Specifically,
when an \ebpf map is allocated,
which takes place
before an \ebpf program is loaded
into the kernel,
\system adds additional metadata
(in the case of software-based isolation,
see~\autoref{sec:implementation:sfi})
or tags the map's
memory region (in the case of
hardware-assisted isolation,
see~\autoref{sec:implementation:mte})
to ensure that the \ebpf program
is allowed to access
only data in the map (\circled{4} in~\autoref{fig:design}).

\subsubsection{\ebpf Program Private Data}
\label{sec:construction:private}
Three types of private data
could exist in an \ebpf program:

\noindgras{\ebpf Stack.}
\system places the entire \ebpf stack
in a sandbox
so that an \ebpf program
operates on an isolated stack
at run time (\circled{1} in~\autoref{fig:design}).

\noindgras{\ebpf Context.}
The context,
similar to function parameters,
refers to
the input passed to an \ebpf program
when it is invoked at a hook point.
The \ebpf framework
provides developers
with abstract data structures
representing
the underlying kernel objects
that different types of \ebpf programs
can access.
These data structures
might also contain
fields
from objects that are
pointed to
by the kernel objects.
On the other hand,
not all fields
of the represented
kernel objects
are included
in the abstract data structures.
For example,
the context
\texttt{\_\_sk\_buff}
contains a subset of
the fields in
the kernel object \texttt{sk\_buff}.
One of its fields,
\texttt{\_\_sk\_buff->ifindex},
corresponds to
\texttt{sk\_buff->dev->ifindex}.
During the verification and compilation
stage of an \ebpf program,
accesses to these abstract data structures
are translated into
direct accesses to
the corresponding kernel objects.
Unfortunately,
this creates opportunities
for an \ebpf program
to access fields of
a kernel object
that are not intended
to be accessed
from its abstract data structure.
To address
this issue,
\system copies
only the fields
specified in the context
to the sandbox
so that the \ebpf program
can access only copied fields
(\circled{3} in~\autoref{fig:design}).

\noindgras{Dynamically-allocated Data.}
\ebpf does not natively support
dynamic memory allocation,
because its verifier
cannot resolve at compile time
the bounds of
dynamically-allocated memory.
With dynamic sandboxing,
\system enables dynamic memory allocation
while ensuring its spatial memory safety at run time.
The heap
(\circled{6} in~\autoref{fig:design})
resides in the sandbox,
whereas its bookkeeping
is isolated from \ebpf programs
in the metadata (\circled{7} in~\autoref{fig:design}).

\system allocates one memory page per sandbox (\circled{B} in \autoref{fig:design}),
since the current \ebpf specification
limits the stack size
to $512$ bytes.
\system's sandbox size can be increased 
without any changes to its design 
if \ebpf's
required stack space increases.
The first half of the page
is dedicated to the metadata,
and the second half is reserved
for sandboxing an \ebpf program's
private data.
The context is placed at the top of the sandbox,
whereas the stack is placed at the bottom %
since it grows ``downwards''
from higher to lower addresses.
The remaining space in the sandbox
is used for dynamic memory allocation.

Upon the invocation of an \ebpf program,
\system prepares its context
by copying to the sandbox
only the fields
of the context object
that will be accessed by the program.
In addition to the security advantage,
this approach also improves run-time performance,
as discussed in~\autoref{sec:evaluation:performance}.
For a context object
with a nested structure
(\eg \texttt{\_\_sk\_buff} contains a pointer to the \texttt{bpf\_sock} object),
if the nested objects
will be accessed by the program,
\system recursively copies
their fields
(only those accessed by the program)
to the dynamically-allocated
memory space in the sandbox.
As such,
\emph{all} pointers in the context
point to addresses within the sandbox's heap.
Upon the program's exit,
\system updates the actual kernel object
if any of its copied fields
(including any nested objects)
are modified in the sandbox.

\subsection{Helper functions}
\label{sec:design:helpers}

\ebpf programs interact with the system
(\eg printing debugging messages
or using \ebpf maps)
through helper functions.
Depending on the type of an \ebpf program
and the privileges held by the user
loading it,
helper functions accessible to
the program vary.
For example, users with
the \texttt{CAP\_PERFMON} privilege
have access to
performance monitoring helper functions.

Helper functions provide a secure, kernel-controlled
mechanism to read or write kernel data structures.
To support their use
in the same way as
the native \ebpf framework,
\system keeps the references
to the original kernel objects
in the metadata
when it prepares the context
upon program invocation (\autoref{sec:construction:private}).
This is necessary,
because
sandboxed \ebpf programs
operate
on the copy of
kernel object fields
in the sandbox,
whereas helper functions
operate directly
on kernel objects.
When an \ebpf program calls a helper function,
\system transparently replaces
the pointer to the sandboxed object
by the pointer to the original kernel object
and
syncs the data
in their corresponding fields
if needed.
This also enables \system
to mitigate
\emph{type confusion vulnerabilities} (\eg CVE-2021-34866~\cite{cve-2021-34866}),
where the pointer passed to a helper function
is not of the type
expected by the function.
This type of vulnerabilities is common
in languages like C
with weak memory safety guarantees.

\section{Sandbox Enforcement Mechanisms}
\label{sec:implementation}
\begin{table}[]
	\centering
	\begin{tabular}{|l|r|}
		\hline
		\textbf{Composition} & \textbf{Lines of Source Code} \\ \hline
		Generic sandboxing code   & 1,361                     \\
		SFI-specific code  & 348                     \\
		MTE-specific code  & 80                     \\ \hline
	\end{tabular}
\caption{The size of \system codebase.}
\label{table:loc}
\end{table}

We describe two sandboxing approaches,
a software-only one
based on address masking (\autoref{sec:implementation:sfi})
and a hardware-assisted one
based on
memory tagging
(\autoref{sec:implementation:mte}),
to enforce isolation 
described in~\autoref{sec:design}.
We also propose an alternative hardware-assisted approach with lower performance overhead,
albeit weaker security guarantees 
(\autoref{sec:implementation:alternative}).
As shown in \autoref{table:loc}, 
76\% of \system's
code
is independent of the enforcement mechanism, 
while only 24\% 
depends on the specific mechanism.

\subsection{Sandbox Management}
\label{sec:implementation:management}

Recall in~\autoref{sec:design:isolation} that
\system's sandbox
protects two types of memory regions,
\ebpf maps
and program private data.
Each memory region is managed by a separate
component in the sandbox,
and each component contains the sandboxed data
accessible by \ebpf programs
and the metadata required to manage the component.
In this section,
we detail the implementation
of the metadata for each component.
The next three sections
discuss how \system's
isolation mechanisms
ensure only sandboxed data
is accessible to \ebpf programs.

\noindgras{\ebpf Maps.}
The metadata are identical in both \system and the original \ebpf framework, except that
\system adds additional masks when running software-based isolation (see~\autoref{sec:implementation:sfi}).
\system modifies
the \ebpf maps' allocation logic
to ensure the alignment of the sandboxed data 
satisfies the alignment requirements
of the isolation mechanisms.

\noindgras{\ebpf Program Private Data.}
The metadata contain the original stack pointer, 
the heap management data, 
and the sandbox management information.
The sandbox management information 
describes the synchronized context objects (see \autoref{sec:construction:private}) and, in the case of software-based isolation, the masks used to enforce isolation.
Similar to \ebpf maps,
\system ensures the alignment requirements are met.
To optimize performance, 
we reuse existing sandboxes
to avoid 
constantly allocating
new ones.
We zero-out sandbox data 
before reusing a sandbox
to prevent leakage
of sensitive data.

\system instruments the prologue and epilogue %
of  an \ebpf program using binary rewriting 
to switch between the sandbox and the kernel.
The prologue switches the stack pointer
to point to the stack in the sandbox.
It also
saves the original stack pointer
in the metadata
to prevent the program
from tampering with it
(\circled{2} in~\autoref{fig:design}).
Subsequently, 
the epilogue restores the original stack pointer 
to switch it back to the kernel stack 
upon program exit.

\subsection{Software-based Isolation}
\label{sec:implementation:sfi}

\system uses 
\emph{address masking},
a software fault isolation (SFI) 
based technique,
to transform any memory address 
into an address 
in a memory region 
specified by a mask~\cite{small1998misfit, koning2017no}.
By masking the \emph{target} addresses
of all load and store instructions
in an \ebpf program,
\system enforces its spatial memory safety 
in such a way that
\emph{all} memory accesses
of the program,
including out-of-bounds accesses (if any), 
always fall within its sandbox.

Since \system's sandbox is separated into
different components for two types of memory regions
(\autoref{sec:design:isolation}),
\system creates a pair of address masks,
\ie an \texttt{and\_mask} and an \texttt{or\_mask},
for each component during sandbox construction,
and stores them 
in the component's 
metadata region.
During
JIT compilation,
\system analyzes an \ebpf program's 
information flow
to differentiate 
between a memory access
to a specific \ebpf map
and to a program instance's private data.
It then inserts
checks through binary rewriting
using the corresponding address masks,
which involves
(1) a bitwise \texttt{and} operation 
between the \texttt{and\_mask} and the target address
to clear its upper bits,
and (2) a bitwise
\texttt{or} operation between the \texttt{or\_mask} 
and the resulting address from (1) 
to map the address to within the sandbox.
\system performs
this binary rewriting
in the last step of the compilation pipeline
(\autoref{fig:design}).
As an example, 
consider a $2048$-byte aligned memory region 
of a sandbox component
located at address \texttt{0xDEADB800}.
\system computes its \texttt{and\_mask} as \texttt{0x7FF}
and \texttt{or\_mask} as \texttt{0xDEADB800}.
If an attacker attempted to 
access out-of-bounds memory 
at \texttt{0xDEAF1234},
address masking would transform the target address 
to \texttt{0xDEADBA34}, 
which would fall in the sandbox.

\subsection{Hardware-assisted Isolation}
\label{sec:implementation:mte}

\begin{figure}[!t]
	\centering
	\includegraphics[width=\columnwidth]{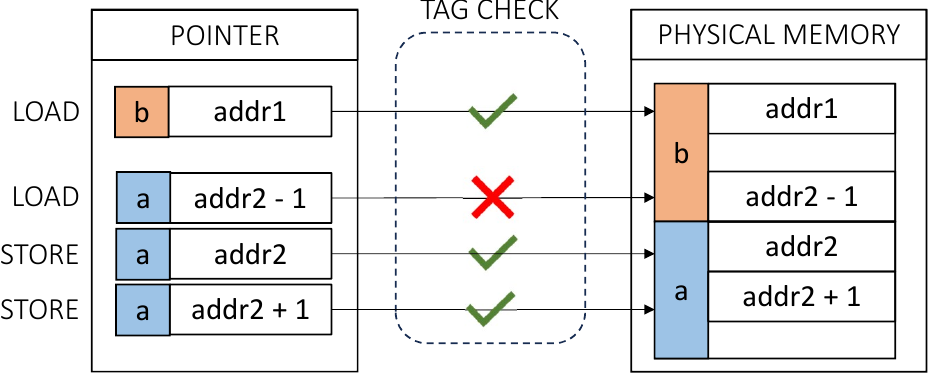}
	\caption{An overview of the MTE mechanism.}
	\label{fig:mte}
\end{figure}

ARM's Memory Tagging Extension (MTE) 
is a hardware primitive 
introduced in the ARMv8.5-A instruction set architecture.
We illustrate how MTE works in \autoref{fig:mte}.
At a high level, 
MTE associates
every 16 bytes of physical memory
with 4-bit metadata
known as a \emph{memory tag}.
It also modifies pointers
to include in each pointer a \emph{pointer tag}
at bits $56$-$59$ of the virtual address.
The MTE-enabled CPU automatically and transparently
checks if the pointer tag 
and the pointed memory's memory tag match on
each load/store operation.
A memory safety violation occurs
when there is a mismatch between the two tags,
and an exception is raised according to one of
the three MTE's modes of operation:
(1) \emph{synchronous} - the kernel raises an exception synchronously upon a mismatch;
(2) \emph{asynchronous} - the kernel 
does not immediately
raise an exception upon a mismatch
but does so
asynchronously;
(3) \emph{asymmetric} - loads are handled in the synchronous mode and stores are handled in the asynchronous mode.

\system configures MTE to operate in the synchronous mode, 
so that spatial memory safety violations %
are immediately detected and 
their impact does not become observerable. %
\system turns on synchronous checks only during \ebpf program execution and restores the original kernel settings 
upon entry/exit of the program 
and \ebpf helper functions.

\system tags the sandbox memory regions
with tag~\tagsandbox~during sandbox construction,
while the rest of the kernel address space 
has a different tag~\tagkernel.
Our tags are consistent with 
Linux's tagging convention~\cite{mte-match-all},
so that 
the kernel can access \ebpf data,
but an \ebpf program
cannot access kernel data outside of its sandbox.

Since MTE tags memory at a 16-byte granularity,
we ensure sandbox memory regions
are 16-byte aligned
with their sizes rounded up to
the nearest multiples of 16 bytes.
\system also tags sandbox pointers with tag~\tagsandbox~so that subsequent sandbox accesses contain 
the same pointer tags as the memory tags.
In contrast to address masking, 
this approach requires no instrumentation of load/store instructions,
because the CPU transparently checks tags.
If an attacker attempted to access memory outside of the sandbox, 
a tag mismatch would occur,
which would raise an exception.
\system would then initiate a kernel panic 
to stop the execution of
the malicious \ebpf program.

In this approach,
\system still \emph{copies} 
context objects that an \ebpf program uses (\autoref{sec:construction:private}).
Alternatively,
we could also leverage the hardware
to avoid the cost
by tagging these objects.
We will explain its trade-offs next.

\subsection{Alternative Hardware-assisted Isolation}
\label{sec:implementation:alternative}

We implement an alternative hardware-assisted approach that does not perform context synchronization
but instead tags (and untags) objects 
(or a subsets of their attribute, 
when appropriate)
accessed by an \ebpf program upon entry/exit. 
Since we guarantee the execution of 
only one \ebpf program \emph{per core}
(see~\autoref{sec:implementation:management}),
if the total number of tags 
is greater than the number of cores,
we might be able to enforce exclusive access 
to tagged objects.
However,
if there are more cores than tags, 
we must reuse tags 
and inevitably weaken
our security guarantees 
(see~\autoref{sec:discussion}
for further discussion on
the implications of limited tags in MTE).
MTE's 16-byte tag granularity
also becomes an issue in this alternative approach.
Rather than tagging
precisely the subset of an object's attributes
that are accessed,
we might have to tag
beyond them
(\eg when an accessed attribute is a 4-byte integer),
or sometimes even beyond an object's boundary
(\eg due to the alignment requirement).
We could address this issue 
by modifying object layouts and alignments,
but the resulting performance and memory implications
would directly contradict our design goals
(see \autoref{sec:design:requirements}).
In \autoref{sec:evaluation:performance}, 
we show that this approach leads to
only minor performance gain;
therefore, it does not justify 
the loss of security guarantees achieved 
in~\autoref{sec:implementation:mte}.

\section{Evaluation}
\label{sec:evaluation}

\begin{figure*}[!t]
	\centering
	\includegraphics[width=\textwidth]{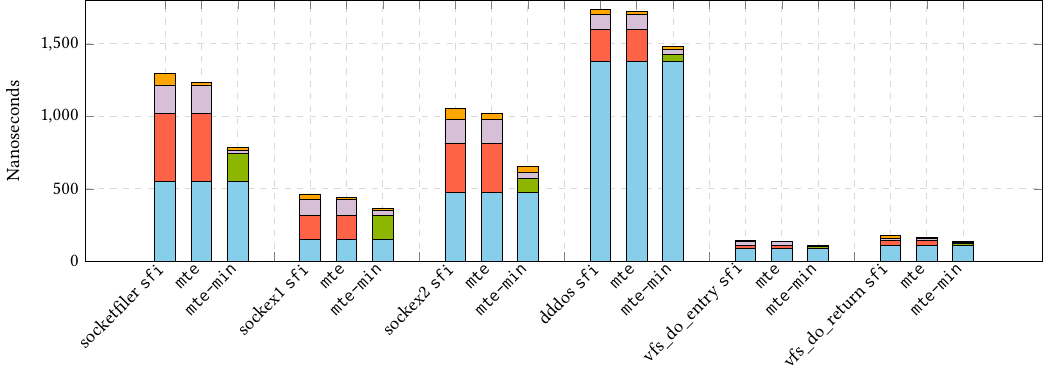}
	\caption{The overhead introduced by \system 
		on six \ebpf programs is divided into \program, \context, \tagging, \sandbox, and \access.}
	\label{fig:cost}
\end{figure*}

We implement \system for the Linux kernel v6.3.8. 
We perform all the evaluation 
on the Fedora Linux Asahi Remix 39 distribution,
on a Mac Mini with an Apple M2 Pro CPU 
with 10 3.5GHz cores, 
16GB of RAM, and a 512GB SSD.
We run experiments on three kernel configurations: 
(1) the \texttt{vanilla} configuration 
runs on the unmodified kernel as our baseline; %
(2) the software-based sandbox configuration 
is denoted by \texttt{sfi}; %
and (3) the hardware-assisted sandbox configuration 
using ARM MTE is denoted by \texttt{mte}
and \texttt{mte-min} 
(for the alternative implementation described 
in~\autoref{sec:implementation:alternative}). %
In this evaluation,
we answer the following three research questions:

\noindgras{Q1.} How does MTE improve
run-time performance of sandboxing
compared to a pure software approach? (see \autoref{sec:evaluation:performance:micro})

\noindgras{Q2.} How much overhead does \system introduce
in realistic workloads? (see \autoref{sec:evaluation:performance:macro})

\noindgras{Q3.} Does \system reduce the attack surface 
introduced by the use of unprivileged \ebpf programs? (see \autoref{sec:evaluation:security})

\subsection{MTE Instruction Analogs}
\label{sec:anologs}

MTE is introduced in ARMv8.5-A,
but it is not supported by any open, widely-available systems 
at the time of this writing.
For example, 
we see no availability of the feature 
on Apple M1 and M2 CPUs (see \autoref{sec:apple}).
Google Pixel 8 supports MTE, but it is a closed system. 
Finally, 
while Amazon's second-generation Neoverse instances
can use MTE,
they are not yet accessible to the general public.
We therefore implemented and tested a \system prototype on QEMU (see \autoref{sec:evaluation:security}).
However, 
running a reliable performance evaluation
on QEMU is hard,
because it does not accurately
represent the clock cycle count
of the actual processor,
even if we use
a supposedly cycle-accurate timing model~\cite{qemu-emulation-model}.
Hence,
to more accurately measure performance overhead, %
we leverage \emph{instruction analogs},
which are also used by prior MTE-related studies~\cite{mckee2022preventing,liljestrand2022color},
to simulate MTE instructions
(\eg \texttt{ldg} and \texttt{stg} for loading and storing tags in memory)
and approximate their CPU cycles and memory footprints.

Memory tagging in \system 
is a one-off operation that takes place 
only when sandboxes are allocated,
either at boot time or on map creation.
Therefore, 
run-time performance overhead stems from 
tag checking operations by the CPU, %
which involve \emph{tag loading} and \emph{tag comparison}.
Similar to prior work~\cite{liljestrand2022color,mckee2022preventing}, we do not simulate the latter,
since it is performed by the hardware~\cite{mtewhitepaper}
and therefore should incur no measurable overhead.
For tag loading,
\system %
emits 
instruction analogs for the tag loading instruction (\texttt{ldg})
upon every \ebpf memory access
to simulate
its cost~\cite{mtewhitepaper}.
Note that these tag loading 
instruction analogs
\emph{overestimate} the cost associated with MTE
by manipulating the 49–53 bits of pointers to simulate address tags and replacing tag loads with regular loads from memory~\cite{mckee2022preventing,liljestrand2022color},
while the actual MTE implementation includes optimizations such as tag caching~\cite{mtewhitepaper}.
Thus,
our evaluation is \emph{conservative},
overestimating \system's overhead rather than underestimating it.

\subsection{Performance}
\label{sec:evaluation:performance}
In~\autoref{sec:evaluation:performance:micro},
we run six
\ebpf programs with varying functionalities 
including network analysis, packet processing, performance tracing, and security,
as a microbenchmark to
measure
the overhead introduced by \system 
in fine granularity.
\autoref{sec:evaluation:performance:macro}
discusses the results of
macrobenchmark experiments
when running realistic server workloads.

\vspace{-0.5pt}
\subsubsection{Microbenchmark}
\label{sec:evaluation:performance:micro}

Based on prior work~\cite{jin2023epf,brunella2022hxdp,hoiland2018express}, 
we select the following programs:

\noindgras{\texttt{socketfiler}}~\cite{libbpf_boostrap}
attaches a socket BPF program to a hook called
\texttt{sock\_queue\-\_rcv\_skb()} to retrieve the protocol and the source and destination IP and port.
This information is written to an \ebpf ring buffer
to be shared with a userspace program.

\noindgras{\texttt{sockex1}}~\cite{linux_bpf_samples}
counts the number of packets associated with a protocol 
observed on an attached network interface.
The program retrieves the protocol attribute of a packet and increments the corresponding entry in an \ebpf map.

\noindgras{\texttt{sockex2}}~\cite{linux_bpf_samples}
performs complex packet parsing.
It uses IP addresses
from
\texttt{struct flow\_keys->dst}
(or hashes of IPv6 \texttt{dst}s) 
to tally the number of packets per IP address.
The program can be attached to the Ethernet interface
to print the statistics every second in userspace.

\noindgras{\texttt{dddos}}~\cite{bcc}
leverages \texttt{kprobe} to track packets arriving at an IP receive entry point (\texttt{ip\_rcv}) 
and monitor the elapsed time between two received packets 
to detect potential denial-of-service attacks.

\noindgras{\texttt{vfs\_do\_entry}} and \textbf{\texttt{vfs\_do\_return}}~\cite{bcc}
attach to the entry and return of the \texttt{vfs\_read} kernel routine, respectively.
These two \ebpf programs combined trace the latency of \texttt{vfs\_read} 
and output it as a histogram distribution 
every five seconds.
We refer to their combined overhead as \texttt{vfs\_read\_lat} in \autoref{tab:netperf} and \autoref{tab:webserver}.

In~\autoref{fig:cost}, we compare the performance overhead
between software-based (\texttt{sfi}), hardware-assisted isolation (\texttt{mte}), and the alternative hardware-assisted implementation described in \autoref{sec:implementation:alternative} (\texttt{mte-min}).
We make the following observations:

\begin{figure}[!t]
	\centering
	\includegraphics[width=\columnwidth]{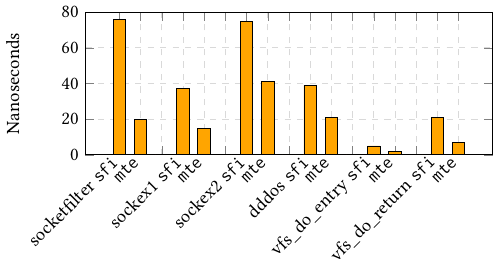}
	\caption{\texttt{sfi} vs \texttt{mte} \access cost. 
		Note that \texttt{mte} and \texttt{mte-min} incur the same cost.}
	\label{fig:sfi-mte}
\end{figure}

\noindent\sandbox overhead
includes the cost of (1) switching between
the original and sandboxed stack,
(2) initializing the metadata
of the heap in the sandbox,
and (3) preparing the mapping
between the original
and the sandboxed context objects,
which is needed for synchronization.
This overhead
is relatively small,
depending neither on the number of load/store operations
nor helper function calls.

\noindent\access overhead
is a function of the number of load/store operations
in an \ebpf program.
Overall, the overhead is relatively small.
\autoref{fig:sfi-mte} zooms in
on it to help visualize the difference between
\texttt{sfi} and \texttt{mte}.
Even with conservative MTE overhead measurement
(in practice, the overhead would be lower),
\texttt{mte} is 45\%-73\%
faster than %
\texttt{sfi}.

\noindent \program measures
the total execution time of the original program,
which in most cases
is dominated by the execution time of helper functions.
\system adds no overhead
\emph{during} the execution of these functions
(although a call to a helper function
requires \system to handle context objects;
this overhead is measured in \context,
as discussed next).
The nature of a helper function influences its execution time.
For example,
logging functions tend to have a longer execution time
than those accessing data.

\begin{figure}[!t]
	\centering
	\includegraphics[width=\columnwidth]{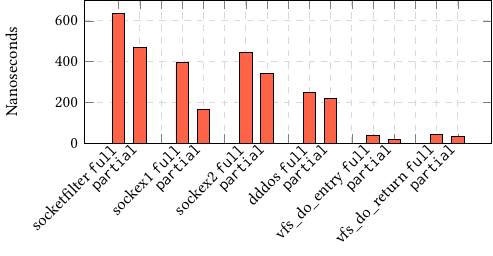}
	\caption{\texttt{full} vs \texttt{partial} \context cost.}
	\label{fig:context}
\end{figure}

\noindent\context
dominates the overall overhead,
which is consistent with the results
from past work~\cite{narayanan2019lxds} on kernel compartmentalization.
This overhead includes
(1) allocating,
at the beginning of execution,
heap memory for context objects
and those recursively
pointed to by these objects;
(2) handling calls to helper functions;
and (3) copying synchronized objects' fields 
in and out of the sandbox during execution.
The overhead depends on the complexity 
of context objects' data structures
(\eg whether they contain nested objects), 
the number of fields accessed in \texttt{read} or \texttt{write} mode, 
and the number helper function calls
and their types.
\autoref{fig:cost} reports the context synchronization cost 
for partial context,
where we copy only fields
accessed by \ebpf programs.
In \autoref{fig:context},
we compare the cost
of synchronizing
\texttt{full} and \texttt{partial} context.
As discussed in~\autoref{sec:construction:private},
synchronizing partial context,
in addition to improving security,
also reduces overhead.
Note that we could have
avoided synchronization almost entirely
in \texttt{mte} (\autoref{sec:implementation:mte}),
but doing so would degrade overall kernel performance
and violate \system's design goal.

\noindent \tagging overhead is incurred only
in the \texttt{mte-min} implementation (see~\autoref{sec:implementation:alternative}) 
when it tags and untags kernel object(s) 
in and out of \ebpf programs. 
This cost is a function of the size of the memory 
that needs to be tagged.
Compared to \texttt{mte},
it reduces the overhead
from two sources: 
(1) \sandbox cost
due to a more simplified sandbox setup,
and (2) \context cost,
as context synchronization
is no longer needed.

\begin{table*}[]
	\centering
	\resizebox{\textwidth}{!}{%
		\begin{tabular}{|lcclccclccc|}
			\hline
			\multicolumn{1}{|c|}{\multirow{2}{*}{Test}} &
			\multicolumn{2}{c|}{\texttt{sockfilter}} &
			\multicolumn{2}{c|}{\texttt{sockex1}} &
			\multicolumn{2}{c|}{\texttt{sockex2}} &
			\multicolumn{2}{c|}{\texttt{dddos}} &
			\multicolumn{2}{c|}{\texttt{vfs\_read\_lat}} \\ \cline{2-11} 
			\multicolumn{1}{|c|}{} &
			\multicolumn{1}{c|}{Vanilla} &
			\multicolumn{1}{c|}{\system MTE} &
			\multicolumn{1}{l|}{Vanilla} &
			\multicolumn{1}{c|}{\system MTE} &
			\multicolumn{1}{c|}{Vanilla} &
			\multicolumn{1}{c|}{\system MTE} &
			\multicolumn{1}{c|}{Vanilla} &
			\multicolumn{1}{c|}{\system MTE} &
			\multicolumn{1}{c|}{Vanilla} &
			\system MTE \\ \hline
			\multicolumn{11}{|c|}{Unidirectional throughput (MB/s)} \\ \hline
			\multicolumn{1}{|l|}{TCP sf} &
			95726 &
			\multicolumn{1}{c|}{88950 (\textbf{7 \%})} &
			\multicolumn{1}{c}{119939} &
			\multicolumn{1}{c|}{115170 (\textbf{4 \%})} &
			114521 &
			\multicolumn{1}{c|}{114486 (\textbf{0 \%})} &
			\multicolumn{1}{c}{100980} &
			\multicolumn{1}{c|}{100741 (\textbf{0 \%})} &
			126057 &
			123267 (\textbf{2 \%}) \\ \cline{1-1}
			\multicolumn{1}{|l|}{TCP c $\rightarrow$ s} &
			54641 &
			\multicolumn{1}{c|}{53270 (\textbf{3 \%})} &
			\multicolumn{1}{c}{75081} &
			\multicolumn{1}{c|}{75507 (\textbf{0 \%})} &
			73581 &
			\multicolumn{1}{c|}{72820 (\textbf{1 \%})} &
			62237 &
			\multicolumn{1}{c|}{61881 (\textbf{1 \%})} &
			80307 &
			80060 (\textbf{0 \%}) \\ \cline{1-1}
			\multicolumn{1}{|l|}{TCP s $\rightarrow$ c} &
			54802 &
			\multicolumn{1}{c|}{53063 (\textbf{3 \%})} &
			81748 &
			\multicolumn{1}{c|}{79079 (\textbf{3 \%})} &
			75840 &
			\multicolumn{1}{c|}{76147 (\textbf{0 \%})} &
			62541 &
			\multicolumn{1}{c|}{61718 (\textbf{1 \%})} &
			80908 &
			80445 (\textbf{1 \%}) \\ \cline{1-1}
			\multicolumn{1}{|l|}{UDP s $\rightarrow$ c} &
			128181 &
			\multicolumn{1}{c|}{121095 (\textbf{6 \%})} &
			138650 &
			\multicolumn{1}{c|}{139525 (\textbf{0 \%})} &
			135468 &
			\multicolumn{1}{c|}{135142 (\textbf{0 \%})} &
			117237 &
			\multicolumn{1}{c|}{118549 (\textbf{0 \%})} &
			144860 &
			140009 (\textbf{3 \%}) \\ \hline
			\multicolumn{11}{|c|}{Round-trip transaction rate (transaction/s)} \\ \hline
			\multicolumn{1}{|l|}{TCP} &
			29851 &
			\multicolumn{1}{c|}{29455 (\textbf{1 \%})} &
			30224 &
			\multicolumn{1}{c|}{29552 (\textbf{2 \%})} &
			30237 &
			\multicolumn{1}{c|}{28973 (\textbf{4 \%})} &
			27426 &
			\multicolumn{1}{c|}{27032 (\textbf{1 \%})} &
			28747 &
			28761 (\textbf{0 \%}) \\ \cline{1-1}
			\multicolumn{1}{|l|}{UDP} &
			32842 &
			\multicolumn{1}{c|}{31429 (\textbf{4 \%})} &
			33369 &
			\multicolumn{1}{c|}{32750 (\textbf{2 \%})} &
			33261 &
			\multicolumn{1}{c|}{31838 (\textbf{4 \%})} &
			29123 &
			\multicolumn{1}{c|}{28679 (\textbf{2 \%})} &
			31013 &
			31379 (\textbf{0 \%}) \\ \hline
		\end{tabular}%
	}
	\caption{Netperf benchmark measuring the overhead of \system on network communications running for 360s. The standard deviations of all results are within 3\% (sf: send file, c $\rightarrow$ s: client to server, s $\rightarrow$ c: server to client).}
	\label{tab:netperf}
\end{table*}     

\vspace{5pt}
\noindent Due to space constraints, 
we report only \texttt{mte} performance results 
in the remaining of the evaluation.
The results of \texttt{sfi} and \texttt{mte-min} 
are available as supplementary material 
(\autoref{sec:sfi} and \autoref{sec:mte-min}).
However, we discuss all implementations in \autoref{sec:evaluation:performance:choice}.

\noindgras{Netperf.}
We use Netperf~\cite{netperf} to measure \system's overhead
on network communications.
The benchmark measures
unidirectional throughputs 
and round-trip latencies for TCP and UDP.
\autoref{tab:netperf} shows that \system introduces only 
0\%-7\% overhead.
This is significantly less than 
what one might
extrapolate from~\autoref{fig:cost},
because
the kernel spends
only a fraction of the total time
in an \ebpf program
when 
sending or receiving network packets.

\subsubsection{Macrobenchmark}
\label{sec:evaluation:performance:macro}

\begin{figure}[!t]
	\centering
	\includegraphics[width=\columnwidth]{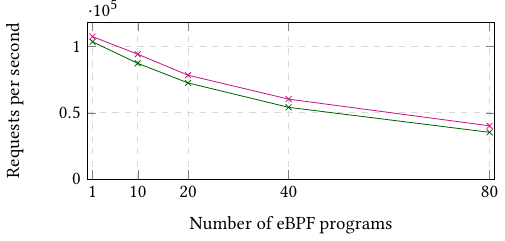}
	\caption{The overhead of \system as a function of the number of \ebpf \texttt{sockex2} programs on
		the \texttt{Apache 200} workload. The standard deviations of all results are within 3\%.
		We report performance results from the \vanilla kernel and from the \system kernel with \mte isolation.}
	\label{fig:scaling}
\end{figure}

\begin{table*}[]
	\centering
	\resizebox{\textwidth}{!}{%
		\begin{tabular}{|lcclccclclc|}
			\hline
			\multicolumn{1}{|c|}{\multirow{2}{*}{Test}} &
			\multicolumn{2}{c|}{\texttt{sockfilter}} &
			\multicolumn{2}{c|}{\texttt{sockex1}} &
			\multicolumn{2}{c|}{\texttt{sockex2}} &
			\multicolumn{2}{c|}{\texttt{dddos}} &
			\multicolumn{2}{c|}{\texttt{vfs\_read\_lat}} \\ \cline{2-11} 
			\multicolumn{1}{|c|}{} &
			\multicolumn{1}{c|}{Vanilla} &
			\multicolumn{1}{c|}{\system MTE} &
			\multicolumn{1}{c|}{Vanilla} &
			\multicolumn{1}{c|}{\system MTE} &
			\multicolumn{1}{c|}{Vanilla} &
			\multicolumn{1}{c|}{\system MTE} &
			\multicolumn{1}{c|}{Vanilla} &
			\multicolumn{1}{c|}{\system MTE} &
			\multicolumn{1}{c|}{Vanilla} &
			\system MTE \\ \hline
			\multicolumn{11}{|c|}{Requests per second (req/s)} \\ \hline
			\multicolumn{1}{|l|}{Apache 100} &
			68808 &
			\multicolumn{1}{c|}{67636 (\textbf{2 \%})} &
			99391 &
			\multicolumn{1}{c|}{95198 (\textbf{4 \%})} &
			97782 &
			\multicolumn{1}{c|}{94858 (\textbf{3 \%})} &
			66868 &
			\multicolumn{1}{c|}{66673 (\textbf{0 \%})} &
			80745 &
			78929 (\textbf{2 \%}) \\
			\multicolumn{1}{|l|}{Apache 200} &
			67862 &
			\multicolumn{1}{c|}{67483 (\textbf{1 \%})} &
			103741 &
			\multicolumn{1}{c|}{100082 (\textbf{4 \%})} &
			101212 &
			\multicolumn{1}{c|}{96699 (\textbf{4 \%})} &
			66770 &
			\multicolumn{1}{c|}{65169 (\textbf{2 \%})} &
			82804 &
			79842 (\textbf{4 \%}) \\
			\multicolumn{1}{|l|}{Apache 500} &
			64510 &
			\multicolumn{1}{c|}{63912 (\textbf{1 \%})} &
			97828 &
			\multicolumn{1}{c|}{97215 (\textbf{1 \%})} &
			95597 &
			\multicolumn{1}{c|}{92692 (\textbf{3 \%})} &
			63170 &
			\multicolumn{1}{c|}{62739 (\textbf{1 \%})} &
			79111 &
			78005 (\textbf{1 \%}) \\
			\multicolumn{1}{|l|}{Apache 1000} &
			63371 &
			\multicolumn{1}{c|}{63496 (\textbf{0 \%})} &
			95188 &
			\multicolumn{1}{c|}{95292 (\textbf{0 \%})} &
			94127 &
			\multicolumn{1}{c|}{91388 (\textbf{3 \%})} &
			64209 &
			\multicolumn{1}{c|}{61715 (\textbf{4 \%})} &
			78547 &
			77203 (\textbf{2 \%}) \\ \hline
			\multicolumn{1}{|l|}{Nginx 100} &
			42376 &
			\multicolumn{1}{c|}{41873 (\textbf{1 \%})} &
			62514 &
			\multicolumn{1}{c|}{61475 (\textbf{2 \%})} &
			53609 &
			\multicolumn{1}{c|}{52543 (\textbf{2 \%})} &
			34615 &
			\multicolumn{1}{c|}{33851 (\textbf{2 \%})} &
			47831 &
			46502 (\textbf{3 \%}) \\
			\multicolumn{1}{|l|}{Nginx 200} &
			42913 &
			\multicolumn{1}{c|}{42569 (\textbf{1 \%})} &
			62481 &
			\multicolumn{1}{c|}{61612 (\textbf{1 \%})} &
			53450 &
			\multicolumn{1}{c|}{52482(\textbf{2 \%})} &
			35050 &
			\multicolumn{1}{c|}{34388 (\textbf{2 \%})} &
			47759 &
			46600 (\textbf{2 \%}) \\
			\multicolumn{1}{|l|}{Nginx 500} &
			42057 &
			\multicolumn{1}{c|}{41813(\textbf{1 \%})} &
			59563 &
			\multicolumn{1}{c|}{58725 (\textbf{1 \%})} &
			51174 &
			\multicolumn{1}{c|}{50454 (\textbf{1 \%})} &
			34895 &
			\multicolumn{1}{c|}{34295 (\textbf{2 \%})} &
			45896 &
			44755 (\textbf{2 \%}) \\
			\multicolumn{1}{|l|}{Nginx 1000} &
			40804 &
			\multicolumn{1}{c|}{40637 (\textbf{0 \%})} &
			55778 &
			\multicolumn{1}{c|}{55061 (\textbf{1 \%})} &
			48063 &
			\multicolumn{1}{c|}{47458 (\textbf{1 \%})} &
			33470 &
			\multicolumn{1}{c|}{33252 (\textbf{1 \%})} &
			43224 &
			42152 (\textbf{2 \%}) \\ \hline
		\end{tabular}%
	}
	\caption{Macrobenchmark measuring web server performance for 100-1000 concurrent connections.}
	\label{tab:webserver}
\end{table*}

In \autoref{fig:cost}, we show the overhead introduced by \system when executing \ebpf programs \emph{alone}.
In absolute values, 
these overheads are less than 685$ns$. 
\ebpf programs are generally triggered during the execution of system calls. 
Their execution time is largely dwarfed by 
the time spent in I/O, context switching, 
and user-space application logic.
Consequently, we expect the overall degradation of 
a user-space application's performance
to be minimal.

We run a set of macrobenchmarks 
from the Phoronix Test Suite~\cite{larabel2011phoronix} 
to measure the overhead introduced by \system on web server workloads (\ie\emph{Apache} and \emph{Nginx}).
The benchmark measures the number of requests/second 
processed with an increasing number of concurrent requests ($100 - 1,000$).
As seen in \autoref{tab:webserver}, \system introduces 0-4\% overhead.

In \autoref{fig:scaling}, we show the number of requests/second as a function of the number of loaded \texttt{sockex2} \ebpf programs on the \texttt{Apache 200} workload from the Phoronix Test Suite.
We select a single program-workload pair due to space constraints.
The selected program-workload pair is the one where we observe the highest overhead in \autoref{tab:webserver} to ensure that our performance reporting is conservative.
\system performance degrades gracefully with 12\% overhead when 80 programs are attached.
In practice, it is unlikely that more than a couple of \ebpf programs will be triggered on the same hook.

\subsubsection{Comparison of Isolation Approaches}
\label{sec:evaluation:performance:choice}

We see 
from \autoref{tab:webserver}, \autoref{sec:sfi}, and \autoref{sec:mte-min} that
the performance difference between the isolation strategies
on macrobenchmark is minimal.
As previously discussed, 
this is explained by the fact that 
the fraction of the time spent 
in executing \ebpf programs 
is small
compared to the overall execution time of typical userspace applications.
MTE-based approaches fault on out-of-bounds accesses, 
unlike the SFI-based one,
which silently restricts memory accesses
within the sandbox.
Faulting is the better behavior 
from a security perspective.
Similarly,
we see minimal practical performance improvement
from the \texttt{mte-min} approach
but weakened security guarantees.
Finally, we emphasize that the overhead
of both MTE-based approaches 
is overestimated 
due to the conservative nature of 
instruction analogs (see \autoref{sec:anologs}).
Consequently, 
we believe \system's \texttt{mte} approach is practical.

\subsection{Security}
\label{sec:evaluation:security}

\begin{table*}[]
	\centering
	\resizebox{0.9\textwidth}{!}{
		\begin{tabular}{|l|p{5.5in}|c|}
			\hline
			\multicolumn{1}{|c|}{\textbf{CVE}} &
			\multicolumn{1}{c|}{\textbf{Vulnerability Description}} &
			\multicolumn{1}{c|}{\textbf{Mitigated}}\\ \hline
			CVE-2023-2163~\cite{cve-2023-2163} &
			Incorrect verifier pruning causes the verifier
                to mark unsafe code paths as safe, which leads to arbitrary read/write in kernel memory, lateral privilege escalation, and container escape. & \textcolor{darkgreen}{\checkmark}\\ 
			\hline

			CVE-2022-23222~\cite{cve-2022-23222} &
            The verifier incorrectly allows
            pointer arithmetic via certain 
            \texttt{*\_OR\_NULL} pointer types,
            which allows out-of-bounds read/write in kernel memory.
            Local users can exploit this vulnerability
            to gain privilege.
            & \textcolor{darkgreen}{\checkmark}\\
			 \hline
			CVE-2021-4204~\cite{cve-2021-4204} &			
			The verifier does not properly validate the bounds of
                inputs to \texttt{bpf\_ringbuf\_submit} and \texttt{bpf\_ringbuf\_discard}, 
            thus allowing out-of-bounds read/write in kernel memory. & \textcolor{darkgreen}{\checkmark}\\
			\hline
			CVE-2021-3490~\cite{cve-2021-3490} &
			The verifier incorrectly tracks the bounds of ALU32 bitwise operations that could lead to out-of-bounds read/write in kernel memory. & \textcolor{darkgreen}{\checkmark}\\ 
			\hline
			CVE-2021-31440~\cite{cve-2021-31440} &
			A bug in the propagation of 32-bit
			unsigned bounds from their 64-bit counterparts 
                in the verifier enables out-of-bounds read/write in kernel memory. & \textcolor{darkgreen}{\checkmark}\\ 
			\hline
			CVE-2020-8835~\cite{cve-2020-8835} &
			The verifier does not properly restrict the register bounds for 32-bit operations, which leads to out-of-bounds read/write in kernel memory.  & \textcolor{darkgreen}{\checkmark}\\ 
			\hline
			CVE-2020-27194~\cite{cve-2020-27194} &
			The verifier mishandles \texttt{scalar32\_min\_max\_or} bounds-tracking during the use of 64-bit values, which allows out-of-bounds read/write in kernel memory. & \textcolor{darkgreen}{\checkmark}\\ 
			\hline
		\end{tabular}
	}
\caption{A list of evaluated vulnerabilities.}
\label{tab:exploit}
\end{table*}

\noindgras{Preventing Known Vulnerabilities.} 
We evaluate %
\system
against seven \emph{high severity} vulnerabilities,
each of which 
has a publicly-available, working proof-of-concept exploit
that leads to arbitrary kernel memory access
and subsequently privilege escalation
(\autoref{tab:exploit}).
To test each exploit,
we backport \system to a kernel version
where the vulnerability is active;
we make the corresponding kernel patches available 
for reproducibility. %
\system successfully prevents \textit{all} exploits.
For other memory access vulnerabilities in~\autoref{sec:background:runtime_safety},
we analytically confirm that 
they would be prevented by \system.
Most of these vulnerabilities
have no proof-of-concept exploits;
developing them from high-level descriptions
and Linux patches %
is non-trivial~\cite{chen2020koobe}.
While we consider such an endeavor useful,
it is beyond the scope of this work.

\smallskip
\noindgras{Fault Injection.}
To further demonstrate \system's effectiveness,
we randomly inject out-of-bounds load and store operations
to the output of the JIT compiler
before \system's final binary rewriting step.
This experiment simulates
illegal memory accesses 
either undetected by the verifier 
or introduced during the JIT compilation.
We repeat fault injections for $10,000$ times,
and \system prevents \emph{all} of them.

\section{Discussion \& Future Work}
\label{sec:discussion}
\noindgras{Performance, Security, and Functionality.}
\system's primary objective is to allow developers
to safely deploy unprivileged \ebpf programs.
The inability to do so
in current Linux distributions
is not only a functionality problem,
but also a security one.
We have witnessed
a plethora of complex workarounds
to run unprivileged programs in privileged mode
by developers
who are motivated by all the
use cases for these programs.
In a way,
disabling unprivileged \ebpf programs
does not improve security,
but on the contrary, increases the potential attack surface.
While \system incurs performance overhead,
it is a concrete step towards resolving this problem.

\noindgras{\texttt{cgroup}s and Memory Tags.}
\label{sec:discussion:tag}
\texttt{cgroup}s are used to organize and limit
resource (\eg CPU and memory) usage
in a hierarchical fashion.
They have gained popularity,
alongside other namespaces,
with the advent of containers.
For example, in Kubernetes,
the \texttt{cgroup} hierarchy is used
to group containers logically.
A number of \ebpf programs, such as socket filters~\cite{cgroup-sock-bpf} and LSM-BPF~\cite{lim2021secure,cgroup-lsm-bpf},
can be associated (and only triggered) within a given \texttt{cgroup}
and its descendants
(see Kubernetes documentation~\cite{kubernetes-ebpf}).
One way to use \system is
to associate sandboxes %
to \texttt{cgroup}s.
In case of MTE,
this would mean associating a specific tag
to each \texttt{cgroup} loading \ebpf programs and maps.
Doing so has two advantages.
First,
it guarantees no data leakage
through \ebpf programs across namespaces.
Example use cases include individual containers safely deploying security audit policies~\cite{lim2021secure}
and scheduling policies~\cite{ghOSt_ebpf, ghOSt_ebpf2,humphries2021ghost}. %
Second,
since MTE supports only 16 different tags,
grouping \ebpf programs and maps by \texttt{cgroup}
would make efficient use of the limited available tags.
Past work~\cite{mckee2022preventing} has suggested
a combination of MTE and ARM's pointer authentication feature~\cite{liljestrand2019pac}
to significantly extend the number of isolation domains
that can be created.
Namespacing certain kernel features and their customization through \ebpf requires careful consideration~\cite{sun2018security,lim2021secure}
beyond the scope of this paper.
We leave the exploration of such an approach to future work and conjecture that it might increase overhead.

\noindgras{Memory Tagging in Other CPU Architectures.}
\label{discussion:tagging}
Features in other CPU architectures, such as Intel PKS~\cite{2021-intel-pks} and memory tagging in lowerRISC~\cite{lowerRISC},
provide functionalities similar to ARM MTE.
We believe that 
the general design proposed in \autoref{sec:design}
should accommodate all of these architectures,
even though the enforcement granularity may vary (\eg PKS provides a 4KB page-size granularity).
We intended to perform a comparison 
between Intel PKS and ARM MTE;
however, issues around PKS availability made us reconsider (see more details in \autoref{sec:pks}).

\noindgras{\ebpf Tokens.}
\label{discussion:token}
Ongoing work~\cite{bpf_token}
proposes to use
\ebpf tokens
to manage
\ebpf program privileges.
Simply put,
\ebpf tokens allow
privileged processes
to delegate privileges
(\eg access to certain \ebpf program types
or helper functions)
to unprivileged processes.
\ebpf tokens are complementary to \system
but orthogonal to the discussion of this paper.

\noindgras{Sandboxing and Speculative Execution.}
\label{discussion:speculative}
As discussed in \autoref{sec:threat}, speculative execution vulnerabilities are out of scope in this paper.
However, we note that sandboxing techniques, similar to those proposed here, have been presented as a potential solution to some speculative execution vulnerabilities~\cite{narayan2021swivel,cauligi2022sok}.
We leave the exploration of this topic to future work.

\noindgras{Software-based Isolation and Silent Failure.}
\label{discussion:failure}
Sandbox enforcement through software-based isolation
currently fails silently,
because all invalid memory accesses are transformed
into accesses inside a sandbox.
Alternatively,
we can add unmmaped guard zones surrounding a sandbox
to increase the likelihood of faults
upon out-of-bounds accesses~\cite{wahbe1993efficient}.

\noindgras{Helper Functions and Access to Kernel Functions.}
\label{discussion:helpers}
As discussed in \autoref{sec:threat},
\system relies on \ebpf helper functions
to perform safe operations.
Indeed, there is no point in sandboxing \ebpf programs,
if attackers can simply use a helper function
to modify credential data structures
and give themselves root access.
Accesses to helper functions
are limited per \ebpf program type.
Helper functions accessible to unprivileged programs
must be vetted extremely carefully.

We notice a growing trend where restrictions imposed on \ebpf programs are being lifted
(\eg \texttt{kfuncs}~\cite{ebpf-kfuncs} in which trusted \ebpf programs can gain ``raw'' access to standard kernel functions).
There is a spectrum of use cases for \ebpf,
from fully flexible customization of the kernel by a trusted party, to allowing a more restricted but safe program to be deployed by arbitrary applications.
While the latter, closer to the original \ebpf vision,
is losing ground due to a large number of vulnerabilities discovered in the past few years, \system can help make it possible.

\section{Related Work}
\label{sec:rw}
\noindgras{In-kernel Sandboxing.}
\system is an in-kernel sandboxing framework.
Unlike \system,
prior in-kernel sandboxing
work~\cite{chou2001empirical,chen2011linux}
focuses on \emph{device drivers},
because they were
(and continue to be)
a major source of bugs 
in the Linux kernel.
However, 
techniques to sandbox device drivers 
are not directly applicable 
to sandboxing \ebpf programs. %
For example, 
microdrivers~\cite{ganapathy2007microdrivers}
and Decaf~\cite{renzelmann2009decaf}
partition a device driver's
code into two components,
migrating one
that contains
non-performance-critical code
to userspace.
As such,
they leverage
the hardware-enforced user/kernel privilege separation
to
``sandbox'' the user-level component
of a driver.
However,
since \ebpf programs are often
placed on critical paths,
constantly accessing kernel data
and calling kernel-level helper functions,
the performance cost from context switching 
would be daunting.
Later work~\cite{nikolaev2013virtuos,narayanan2019lxds,narayanan2020lightweight} proposes to
sandbox a driver
in a separate virtual address space,
but using similar techniques
on \ebpf programs
would again incur high cost
of switching 
between virtualization domains.

\system's software-based isolation approach
has been used to sandbox 
device drivers~\cite{erlingsson2006xfi,mao2011software},
but prior work
often omits to check read instructions 
due to performance concerns~\cite{erlingsson2006xfi,mao2011software}.
\system %
instruments all loads and stores %
to ensure both confidentiality and integrity 
of kernel memory.

HAKC~\cite{mckee2022preventing} 
uses a combination of ARM's MTE and Pointer Authentication 
to sandbox device drivers,
but it requires user annotations 
to specify isolation policy 
and marshal data across isolated compartments.
\system instruments \ebpf programs
fully automatically,
providing desirable security gurantees
without increasing the burden
for developers.

\noindgras{\ebpf Sandboxing.}
\system 
is built upon SandBPF~\cite{lim2023ebpf},
with numerous extensions and improvements,
including
(1) introducing hardware-assisted isolation
while SandBPF is only software-based;
(2) simplifying access checking 
and
thus reducing the number of  
additional instructions to be executed; 
(3) reducing context synchronization overhead 
by copying only data fields
accessed by \ebpf programs
while SandBPF copies entire context objects;
(4) protecting against type confusion vulnerabilities
when calling helper functions while SandBPF does not;
(5) supporting \emph{any} \ebpf programs,
while SandBPF cannot run programs
that use \ebpf maps;
and
(6) being compatible with both x86 and ARM architectures.

\noindgras{Improving \ebpf Security.}
Improving the security of \ebpf programs 
is an active area of research.
Using formal techniques
to improve the quality of
the built-in verifier~\cite{vishwanathan2023verifying,vishwanathan2022sound,bhat2022formal}
is one important line of work,
while using fuzzing techniques~\cite{mohamed2023understanding,hung2023brf,li2023fuzzing,fuzzing-ebpf-1,fuzzing-ebpf-2}
to detect vulnerabilities in the \ebpf framework
is another.
Jia~\etal~\cite{jia2023kernel} recently 
proposed to move verification out of the kernel 
by leveraging Rust 
and a cryptographic signature scheme.
These techniques are likely insufficient on their own,
but they are complementary to \system,
which provides defense-in-depth 
with a relatively low overhead.

\section{Conclusion}
\label{sec:conclusion}
We show that 
dynamic sandboxing improves memory safety
of \ebpf programs,
complementary to the static mechanism
employed by the \ebpf verifier.
While we do not foresee it 
replacing verification,
it enhances the kernel's run-time safety,
particularly when running unprivileged \ebpf programs.
Our framework, \system,
implements dynamic sandboxing
using both software-based
and hardware-assisted approaches,
imposing minimal overhead %
to make its adoption practical.

\begin{acks}
We thank ATC 2024, EuroSys 2024 and CCSW 2024 reviewers for their help in improving the paper.
We acknowledge the support of the Natural Sciences and Engineering Research Council of Canada (NSERC). Nous remercions le Conseil de
recherches en sciences naturelles et en génie du Canada (CRSNG) de son soutien.
This work was supported by Mitacs through the Mitacs Globalink Research Internship and the Globalink Graduate Fellowship programs.
Cette recherche a reçu le soutien de Mitacs dans le cadre des programmes Stage de recherche Mitacs Globalink et Bourse aux cycles supérieurs Globalink.

\end{acks}

\bibliographystyle{ACM-Reference-Format}
\bibliography{bibliography}
\balance

\appendix

\section*{Supplementary Material}

\section{SFI Results}
\label{sec:sfi}
As shown in~\autoref{tab:webserver_sfi}, our \texttt{sfi} implementation incurs 0\%-4\% overhead on the Apache and Nginx webserver benchmarks.

\begin{table*}[]
	\centering
	\resizebox{\textwidth}{!}{%
		\begin{tabular}{|lcclccclclc|}
			\hline
			\multicolumn{1}{|c|}{\multirow{2}{*}{Test}} &
			\multicolumn{2}{c|}{\texttt{sockfilter}} &
			\multicolumn{2}{c|}{\texttt{sockex1}} &
			\multicolumn{2}{c|}{\texttt{sockex2}} &
			\multicolumn{2}{c|}{\texttt{dddos}} &
			\multicolumn{2}{c|}{\texttt{vfs\_read\_lat}} \\ \cline{2-11} 
			\multicolumn{1}{|c|}{} &
			\multicolumn{1}{c|}{Vanilla} &
			\multicolumn{1}{c|}{\system SFI} &
			\multicolumn{1}{c|}{Vanilla} &
			\multicolumn{1}{c|}{\system SFI} &
			\multicolumn{1}{c|}{Vanilla} &
			\multicolumn{1}{c|}{\system SFI} &
			\multicolumn{1}{c|}{Vanilla} &
			\multicolumn{1}{c|}{\system SFI} &
			\multicolumn{1}{c|}{Vanilla} &
			\system SFI \\ \hline
			\multicolumn{11}{|c|}{Requests per second (req/s)} \\ \hline
			\multicolumn{1}{|l|}{Apache 100} &
			68808 &
			\multicolumn{1}{c|}{67497 (\textbf{2 \%})} &
			99391 &
			\multicolumn{1}{c|}{96195 (\textbf{3 \%})} &
			97782 &
			\multicolumn{1}{c|}{95619 (\textbf{2 \%})} &
			66868 &
			\multicolumn{1}{c|}{64894 (\textbf{3 \%})} &
			80745 &
			79346 (\textbf{2 \%}) \\
			\multicolumn{1}{|l|}{Apache 200} &
			67862 &
			\multicolumn{1}{c|}{67664 (\textbf{0 \%})} &
			103741 &
			\multicolumn{1}{c|}{103063 (\textbf{1 \%})} &
			101212 &
			\multicolumn{1}{c|}{100476 (\textbf{1 \%})} &
			66770 &
			\multicolumn{1}{c|}{66936 (\textbf{0 \%})} &
			82804 &
			79872 (\textbf{4 \%}) \\
			\multicolumn{1}{|l|}{Apache 500} &
			64510 &
			\multicolumn{1}{c|}{64035 (\textbf{1 \%})} &
			97828 &
			\multicolumn{1}{c|}{97214 (\textbf{1 \%})} &
			95597 &
			\multicolumn{1}{c|}{93063 (\textbf{3 \%})} &
			63170 &
			\multicolumn{1}{c|}{60661 (\textbf{4 \%})} &
			79111 &
			77866 (\textbf{2 \%}) \\
			\multicolumn{1}{|l|}{Apache 1000} &
			63371 &
			\multicolumn{1}{c|}{62250 (\textbf{2 \%})} &
			95188 &
			\multicolumn{1}{c|}{95482 (\textbf{0 \%})} &
			94127 &
			\multicolumn{1}{c|}{91378 (\textbf{3 \%})} &
			64209 &
			\multicolumn{1}{c|}{62482 (\textbf{3 \%})} &
			78547 &
			76197 (\textbf{3 \%}) \\ \hline
			\multicolumn{1}{|l|}{Nginx 100} &
			42376 &
			\multicolumn{1}{c|}{41519 (\textbf{2 \%})} &
			62514 &
			\multicolumn{1}{c|}{61489 (\textbf{2 \%})} &
			53609 &
			\multicolumn{1}{c|}{52163 (\textbf{3 \%})} &
			34615 &
			\multicolumn{1}{c|}{34325 (\textbf{1 \%})} &
			47831 &
			46404 (\textbf{3 \%}) \\
			\multicolumn{1}{|l|}{Nginx 200} &
			42913 &
			\multicolumn{1}{c|}{42431 (\textbf{1 \%})} &
			62481 &
			\multicolumn{1}{c|}{61548 (\textbf{1 \%})} &
			53450 &
			\multicolumn{1}{c|}{52068 (\textbf{3 \%})} &
			35050 &
			\multicolumn{1}{c|}{34964 (\textbf{0 \%})} &
			47759 &
			46015 (\textbf{4 \%}) \\
			\multicolumn{1}{|l|}{Nginx 500} &
			42057 &
			\multicolumn{1}{c|}{41523 (\textbf{1 \%})} &
			59563 &
			\multicolumn{1}{c|}{58612 (\textbf{2 \%})} &
			51174 &
			\multicolumn{1}{c|}{50081 (\textbf{2 \%})} &
			34895 &
			\multicolumn{1}{c|}{34557 (\textbf{1 \%})} &
			45896 &
			44439 (\textbf{3 \%}) \\
			\multicolumn{1}{|l|}{Nginx 1000} &
			40804 &
			\multicolumn{1}{c|}{40546 (\textbf{1 \%})} &
			55778 &
			\multicolumn{1}{c|}{54907 (\textbf{2 \%})} &
			48063 &
			\multicolumn{1}{c|}{46998 (\textbf{2 \%})} &
			33470 &
			\multicolumn{1}{c|}{33123 (\textbf{1 \%})} &
			43224 &
			41876 (\textbf{3 \%}) \\ \hline
		\end{tabular}%
	}
	\caption{Macrobenchmark measuring web server performance of \system SFI for 100-1000 concurrent connections.}
	\label{tab:webserver_sfi}
\end{table*}

\section{MTE-min Results}
\label{sec:mte-min}
As shown in~\autoref{tab:webserver_mte_min}, our \texttt{mte-min} implementation incurs 0\%-4\% overhead on the Apache and Nginx webserver benchmarks.

\begin{table*}[]
	\centering
	\resizebox{\textwidth}{!}{%
		\begin{tabular}{|lcclccclclc|}
			\hline
			\multicolumn{1}{|c|}{\multirow{2}{*}{Test}} &
			\multicolumn{2}{c|}{\texttt{sockfilter}} &
			\multicolumn{2}{c|}{\texttt{sockex1}} &
			\multicolumn{2}{c|}{\texttt{sockex2}} &
			\multicolumn{2}{c|}{\texttt{dddos}} &
			\multicolumn{2}{c|}{\texttt{vfs\_read\_lat}} \\ \cline{2-11} 
			\multicolumn{1}{|c|}{} &
			\multicolumn{1}{c|}{Vanilla} &
			\multicolumn{1}{c|}{\system MTE-min} &
			\multicolumn{1}{c|}{Vanilla} &
			\multicolumn{1}{c|}{\system MTE-min} &
			\multicolumn{1}{c|}{Vanilla} &
			\multicolumn{1}{c|}{\system MTE-min} &
			\multicolumn{1}{c|}{Vanilla} &
			\multicolumn{1}{c|}{\system MTE-min} &
			\multicolumn{1}{c|}{Vanilla} &
			\system MTE-min \\ \hline
			\multicolumn{11}{|c|}{Requests per second (req/s)} \\ \hline
			\multicolumn{1}{|l|}{Apache 100} &
			68808 &
			\multicolumn{1}{c|}{68645 (\textbf{0 \%})} &
			99391 &
			\multicolumn{1}{c|}{98758 (\textbf{1 \%})} &
			97782 &
			\multicolumn{1}{c|}{95907 (\textbf{2 \%})} &
			66868 &
			\multicolumn{1}{c|}{66314 (\textbf{1 \%})} &
			80745 &
			79037 (\textbf{2 \%}) \\
			\multicolumn{1}{|l|}{Apache 200} &
			67862 &
			\multicolumn{1}{c|}{67395 (\textbf{1 \%})} &
			103741 &
			\multicolumn{1}{c|}{99957 (\textbf{4 \%})} &
			101212 &
			\multicolumn{1}{c|}{98406 (\textbf{3 \%})} &
			66770 &
			\multicolumn{1}{c|}{65202 (\textbf{2 \%})} &
			82804 &
			79549 (\textbf{4 \%}) \\
			\multicolumn{1}{|l|}{Apache 500} &
			64510 &
			\multicolumn{1}{c|}{64827 (\textbf{0 \%})} &
			97828 &
			\multicolumn{1}{c|}{96679 (\textbf{1 \%})} &
			95597 &
			\multicolumn{1}{c|}{94498 (\textbf{1 \%})} &
			63170 &
			\multicolumn{1}{c|}{62743 (\textbf{1 \%})} &
			79111 &
			76833 (\textbf{3 \%}) \\
			\multicolumn{1}{|l|}{Apache 1000} &
			63371 &
			\multicolumn{1}{c|}{63322 (\textbf{0 \%})} &
			95188 &
			\multicolumn{1}{c|}{94044 (\textbf{1 \%})} &
			94127 &
			\multicolumn{1}{c|}{91643 (\textbf{3 \%})} &
			64209 &
			\multicolumn{1}{c|}{63582 (\textbf{1 \%})} &
			78547 &
			75960 (\textbf{3 \%}) \\ \hline
			\multicolumn{1}{|l|}{Nginx 100} &
			42376 &
			\multicolumn{1}{c|}{41571 (\textbf{2 \%})} &
			62514 &
			\multicolumn{1}{c|}{61148 (\textbf{2 \%})} &
			53609 &
			\multicolumn{1}{c|}{52758 (\textbf{2 \%})} &
			34615 &
			\multicolumn{1}{c|}{34109 (\textbf{1 \%})} &
			47831 &
			46604 (\textbf{3 \%}) \\
			\multicolumn{1}{|l|}{Nginx 200} &
			42913 &
			\multicolumn{1}{c|}{42389 (\textbf{1 \%})} &
			62481 &
			\multicolumn{1}{c|}{61251 (\textbf{2 \%})} &
			53450 &
			\multicolumn{1}{c|}{52700 (\textbf{1 \%})} &
			35050 &
			\multicolumn{1}{c|}{34482 (\textbf{2 \%})} &
			47759 &
			46549 (\textbf{3 \%}) \\
			\multicolumn{1}{|l|}{Nginx 500} &
			42057 &
			\multicolumn{1}{c|}{42046 (\textbf{0 \%})} &
			59563 &
			\multicolumn{1}{c|}{58651 (\textbf{2 \%})} &
			51174 &
			\multicolumn{1}{c|}{50740 (\textbf{1 \%})} &
			34895 &
			\multicolumn{1}{c|}{34596 (\textbf{1 \%})} &
			45896 &
			44737 (\textbf{3 \%}) \\
			\multicolumn{1}{|l|}{Nginx 1000} &
			40804 &
			\multicolumn{1}{c|}{40722 (\textbf{0 \%})} &
			55778 &
			\multicolumn{1}{c|}{54842 (\textbf{2 \%})} &
			48063 &
			\multicolumn{1}{c|}{47612 (\textbf{1 \%})} &
			33470 &
			\multicolumn{1}{c|}{33122 (\textbf{1 \%})} &
			43224 &
			42223 (\textbf{2 \%}) \\ \hline
		\end{tabular}%
	}
	\caption{Macrobenchmark measuring web server performance of \system MTE-min for 100-1000 concurrent connections.}
	\label{tab:webserver_mte_min}
\end{table*}

\section{Modifications to Test Programs}
\label{sec:modification}
We modified \texttt{sockex1} and \texttt{sockex2} 
to use the helper function called 
\texttt{bpf\_skb\_load\_bytes},
instead of the LLVM builtin functions 
(\eg \texttt{load\_bytes}, \texttt{load\_half}, 
and \texttt{load\_word})
to access network packets.
All modified test programs will be made publicly available.

\section{PKS Availability}
\label{sec:pks}
We tested the presence of the PKS feature on a $13^{th}$ generation CPU (i9-13900K) and found that it was not supported.
We checked Intel documentation~\cite{intel-manual},
which states that the feature is present in ``$12^{th}$ generation Intel Core processor based on Alder Lake performance hybrid architecture [and] $4^{th}$
generation Intel Xeon Scalable Processor Family based on Sapphire Rapids microarchitecture''.
We contacted Intel and exchanged numerous e-mails and messages across several months and multiple channels.
We were told that PKS is not supported in the $13^{th}$  generation and that ``the technological feature has been removed from all core processors and Xeon products''.
We also note that patches to bring PKS support to the Linux~\cite{intel-pks-patch,intel-pks-patch-v10} have not been merged.
The only mention in the Linux code base 
is the definition of \texttt{X86\_FEATURE\_PKS} as of kernel release 6.7-rc5.
We could not find any official announcement.
We refer interested readers to the work by Lu \etal~\cite{lu2023moat}.

\section{MTE Support on Apple M1 and M2}
\label{sec:apple}
Some sources on the internet state that Apple M1 and M2 CPUs are based on ARMv8.5A, but it does not appear to be the case or at least not all features are available or supported.
First, we relied on Linux kernel CPU feature check~\cite{arm-feature-boot-check} performed at boot time.
MTE did not appear as an available feature.
We then followed ARM instructions~\cite{arm-feature-check} to perform a sanity check, and confirmed that the feature was indeed unavailable.
At the time of this writing, it is not clear if there is an open platform supporting MTE.
Multiple online discussions seem to indicate that the latest Google Pixel 8 should support MTE, but the platform is closed.
We are expecting accessibility to MTE to improve over the next few months or perhaps years.
For example, 
Amazon has announced that its second-generation Neoverse CPU will support MTE.
However, access to Neoverse V2 instances are currently restricted.
We asked for early access,
but our request was rejected at the time of this submission.

\end{document}